\documentclass[aps,prl,twocolumn,superscriptaddress,floatfix]{revtex4-1}

\usepackage[latin9]{inputenc}
\usepackage{graphicx}
\usepackage{dcolumn}
\usepackage{bm}
\usepackage{amssymb}
\usepackage{microtype}
\usepackage{xfrac}
\usepackage{makecell}
\usepackage{array}
\usepackage[version=4]{mhchem}
\usepackage{gensymb}
\usepackage{multirow}
\usepackage[colorlinks=true]{hyperref}

\begin{document}

\title{Charge Density Fluctuations with Enhanced Superconductivity at the Proposed Nematic Quantum Critical Point}

\author{Y.~Chen}
\affiliation{Department of Physics, University of California, Berkeley, CA 94720, USA}
\affiliation{Material Sciences Division, Lawrence Berkeley National Lab, Berkeley, California 94720, USA}

\author{N.~Giles-Donovan}
\affiliation{Department of Physics, University of California, Berkeley, CA 94720, USA}
\affiliation{Material Sciences Division, Lawrence Berkeley National Lab, Berkeley, California 94720, USA}

\author{J.~Guo}
\affiliation{Center for Correlated Matter and School of Physics, Zhejiang University, Hangzhou 310058, China}

\author{R.~Chen}
\affiliation{Center for Correlated Matter and School of Physics, Zhejiang University, Hangzhou 310058, China}

\author{H.~Fukui}
\affiliation{Precision Spectroscopy Division, CSRR, SPring-8/JASRI, 1-1-1 Kouto, Sayo, Hyogo, 679-5198, Japan}

\author{T.~Manjo}
\affiliation{Precision Spectroscopy Division, CSRR, SPring-8/JASRI, 1-1-1 Kouto, Sayo, Hyogo, 679-5198, Japan}

\author{D.~Ishikawa}
\affiliation{Precision Spectroscopy Division, CSRR, SPring-8/JASRI, 1-1-1 Kouto, Sayo, Hyogo, 679-5198, Japan}
\affiliation{Materials Dynamics Laboratory, RIKEN SPring-8 Center, Sayo, Hyogo 679-5148, Japan}

\author{A.~Q.~R.~Baron}
\affiliation{Materials Dynamics Laboratory, RIKEN SPring-8 Center, Sayo, Hyogo 679-5148, Japan}
\affiliation{Precision Spectroscopy Division, CSRR, SPring-8/JASRI, 1-1-1 Kouto, Sayo, Hyogo, 679-5198, Japan}

\author{Y.~Song}
\affiliation{Center for Correlated Matter and School of Physics, Zhejiang University, Hangzhou 310058, China}

\author{R.~J.~Birgeneau}
\affiliation{Department of Physics, University of California, Berkeley, CA 94720, USA}
\affiliation{Material Sciences Division, Lawrence Berkeley National Lab, Berkeley, California 94720, USA}

\date{\today}
\begin{abstract} A quantum critical point (QCP) represents a continuous phase transition at absolute zero. At the QCP of an unconventional superconductor, enhanced superconducting transition temperature and magnetic fluctuations strength are often observed together, indicating magnetism-mediated superconductivity. This raises the question of whether quantum fluctuations in other degrees of freedom, such as charge, could similarly boost superconductivity. However, because charge is frequently intertwined with magnetism, isolating and understanding its specific role in Cooper pair formation poses a significant challenge. Here, we report persistent charge density fluctuations (CDF) down to 15 K in the non-magnetic superconductor Sr$_{0.77}$Ba$_{0.23}$Ni$_{2}$As$_{2}$, which lies near a proposed nematic QCP associated with a six-fold enhancement of superconductivity. Our results show that the quasi-elastic CDF does not condense into resolution-limited Bragg peaks and displays non-saturated strength. The CDF completely softens at 25 K, with its critical behavior described by the same mathematical framework as the antiferromagnetic Fermi liquid model, yielding a fitted Curie-Weiss temperature of $\theta \approx 0$ K. Additionally, we find that the nematic fluctuations are not lattice-driven, as evidenced by the absence of softening in nematic-coupled in-plane transverse acoustic phonons. Our discovery positions Sr$_{x}$Ba$_{1-x}$Ni$_{2}$As$_{2}$ as a promising candidate for charge-fluctuation-driven nematicity and superconductivity.

\end{abstract}

\maketitle
High-temperature superconductivity in both cuprates and iron-based materials originates from a complex landscape of electronic states involving the spin, charge and orbital degrees freedom~\cite{RevModPhys.75.1201,Paglione2010,RevModPhys.83.1589}. While spin fluctuations are commonly regarded as the primary driver of Cooper pair formation in both families~\cite{PhysRevB.57.6165,PhysRevB.38.6614,PhysRevLett.68.1414,ROSSATMIGNOD199186,PhysRevLett.71.919,PhysRevLett.77.5425,PhysRevB.46.5561,Christianson2008,Chen2019}, theoretical frameworks have been developed showing that orbital and charge degrees of freedom can also facilitate superconductivity~\cite{PhysRevLett.114.097001,doi:10.1073/pnas.1620651114,PhysRevLett.103.177001}. However, these non-magnetic degrees of freedom, particularly charge density fluctuations (CDF), are typically regarded as competitors to superconductivity. This view arises from the fact that the condensation of both orders depends on electron fluctuations at the Fermi surface~\cite{annurev:/content/journals/10.1146/annurev-conmatphys-032922-094430}. Interestingly, in iron pnictides, theory has shown that CDF can lead to similarly rich physics involving orbital order and superconductivity just like spin fluctuations~\cite{Fernandes2014}. Thus, it is natural to search for a tunable pnictide or chalcogenide series with a quantum critical point (QCP) associated with a charge density wave (CDW) to enhance the coupling of Cooper pairs. Such studies face significant challenges due to the pervasive magnetic fluctuations, which complicate our understanding of the interplay between pairing and non-magnetic degrees of freedom. In the notable case of the FeSe$_{1-x}$S$_{x}$ series, where coupling between Cooper pairs is considered to be dominated by nematic fluctuations beyond the QCP at \( x_{c} = 0.17 \)~\cite{doi:10.1126/sciadv.aar6419}, the high-energy spin fluctuations are detected by resonant inelastic X-ray scattering techniques~\cite{PhysRevLett.132.016501}. 

The discovery of the nonmagnetic superconductor Sr\(_{x}\)Ba\(_{1-x}\)Ni\(_{2}\)As\(_{2}\) provides a new approach~\cite{Ronning_2008,PhysRevB.78.172504}. This series of materials is isomorphic to the iron pnictide BaFe\(_{2}\)As\(_{2}\), and the parent compound BaNi\(_{2}\)As\(_{2}\) experiences a very subtle tetragonal-orthorhombic structural distortion at 150~K~\cite{PhysRevB.104.184509}. Yet, in BaNi\(_{2}\)As\(_{2}\), an incommensurate CDW (ICDW) instead of a spin density wave (SDW) emerges almost concurrently with the structural distortion~\cite{PhysRevLett.122.147601,PhysRevLett.127.027602}. Notably, signatures of electronic nematicity were observed in BaNi\(_{2}\)As\(_{2}\) and a charge density fluctuation origin was proposed as an analogy to the spin fluctuation scenario in BaFe\(_{2}\)As\(_{2}\)~\cite{PhysRevB.104.184509,Yao2022}. However, experimental difficulties in studying the interplay between ICDW, electronic nematicity, and superconductivity arise from the emergence of the triclinic phase at $T_{s}\approx$135 K, which wipes out the ICDW and signatures of electronic nematicity.

\begin{figure}[t]
    \includegraphics[width=\columnwidth]{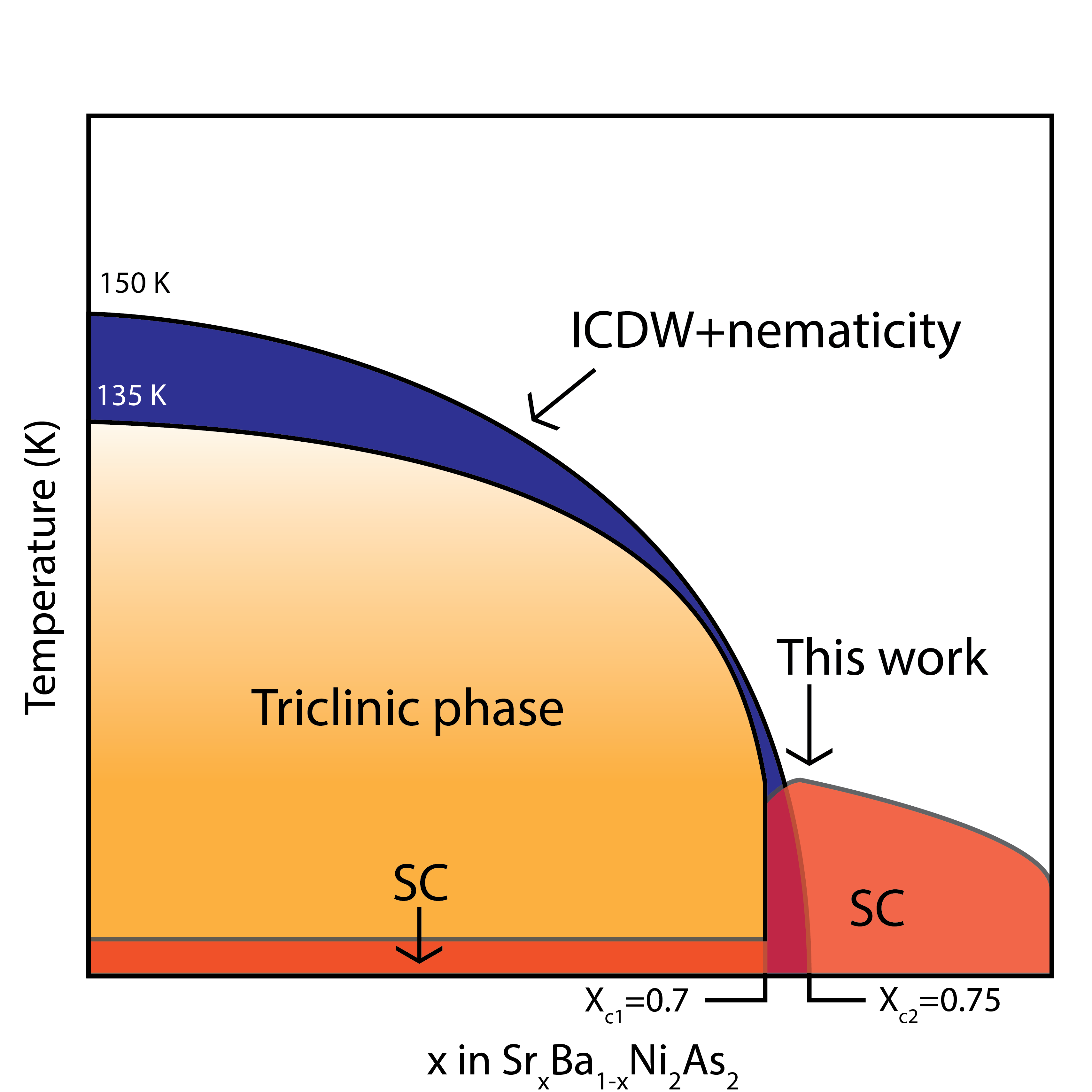}
    \centering
    \caption{Modified sketch of phase diagram of Sr\(_{x}\)Ba\(_{1-x}\)Ni\(_{2}\)As\(_{2}\), based on the phase diagram presented in~\cite{Eckberg2020}. The phase boundary of ICDW is modified with a proposed ICDW QCP. The superconducting transition temperature is exaggerated by 15 times in order to more clearly illustrate the doping dependence. The commensurate charge density wave phases are not shown here as they only exists in the triclinic phase and are irrelevant to our work.}
    \label{Fig_1}
\end{figure}

The complexity introduced by the triclinic phase is lifted by doping Sr onto Ba sites in Sr\(_{x}\)Ba\(_{1-x}\)Ni\(_{2}\)As\(_{2}\), accompanied with a continuously lowered structural transition temperature \(T_{s} \). As shown in the phase diagram presented in Fig.~\ref{Fig_1}, a strong first order quantum phase transition occurs at $x_{c1}$~=~0.7, where a sudden suppression of the triclinc phase and a six-fold enhancement of the bulk superconducting transition temperature (from 0.6 K to 3.5 K) was observed~\cite{Eckberg2020}. Interestingly, the elastoresistivity peak is continuous over the entire doping region until it is fully suppressed around the proposed QCP at \( x_{c2} \approx 0.75 \). This suggests the lattice dynamics that facilitate the triclinic phase are decoupled from the nematicity. On the contrary, the sudden enhancement of bulk \( T_c \) across \( x_{c1} \) and the presence of substantial nematic fluctuations indicates the introduction of a distinct superconductivity where quantum fluctuations are allowed to play a significant role in the formation of Cooper pairs as opposed to the triclinic lattice-mediated superconductivity below \( x_{c1} \) with no noticeable change in bulk \( T_c \) for \( x < 0.7 \). This is further supported by the suppression of $T_{c}$ at higher Sr content, where nematic fluctuations are also weakened, leading to a dome of SC around \( x_{c1} \) or \( x_{c2} \). 

The above brief review shows, after excluding the triclinic phase, the phase diagram of Sr\(_{1-x}\)Ba\(_{x}\)Ni\(_{2}\)As\(_{2}\) resembles that of the doped BaFe\(_{2}\)As\(_{2}\) series, where the ICDW replaces the role of spin density waves. However, critical information regarding CDF near the proposed QCP remains elusive, particularly due to earlier reports that shows the absence of ICDW in materials with higher Sr content (\(x > 0.65\))\cite{PhysRevLett.127.027602,PhysRevB.106.054107}. In this work, we recover the missing information using inelastic X-ray scattering (IXS) techniques. Our findings reveal persistent, non-saturated incommensurate charge density wave (ICDW) fluctuations down to 15 K in optimally doped Sr\(_{0.77}\)Ba\(_{0.23}\)Ni\(_{2}\)As\(_{2}\). This suggests that superconductivity in Sr\(_{0.77}\)Ba\(_{0.23}\)Ni\(_{2}\)As\(_{2}\) may be driven by CDF, offering insight into the role of non-magnetic degrees of freedom in enhancing superconducting properties near the QCP.

All data in this work were taken on BL35XU at SPring-8~\cite{BARON2000461}. We used slit sizes of 15~$\times$~15~mm$^2$ (giving a \textbf{Q} resolution of $\sim$0.02 r.l.u.) and the beamspot was 85~$\mu$m~(horizontal)~$\times$~50~$\mu$m~(vertical) at the sample. The sample was a 1~$\times$~2~mm$^2$ plate-like single crystal mounted on a copper pin with silver epoxy. The sample thickness was $\sim$100$~\mu$m which is comparable to the attenuation length in Sr\(_{0.77}\)Ba\(_{0.23}\)Ni\(_{2}\)As\(_{2}\) at an incident X-ray energy of 21 keV and allowed the experiment to be performed in transmission geometry. The energy resolution on BL35XU was measured using a piece of TEMPAX borosilicate glass according to Ref.~\onlinecite{Ishikawa:ay5575}. This resulted in a full width at half maximum (FWHM) of $\sim$~1.8~meV for the measured instrument energy resolution function. This high energy resolution indicates that elastic scans only involve integrated dynamical structural factors within a 20~K range. As such, the elastic scans reported here should be understood as energy-resolved which offers significant advantages over home-lab single crystal X-ray diffraction experiments.

\begin{figure}[t]
    \includegraphics[width=\columnwidth]{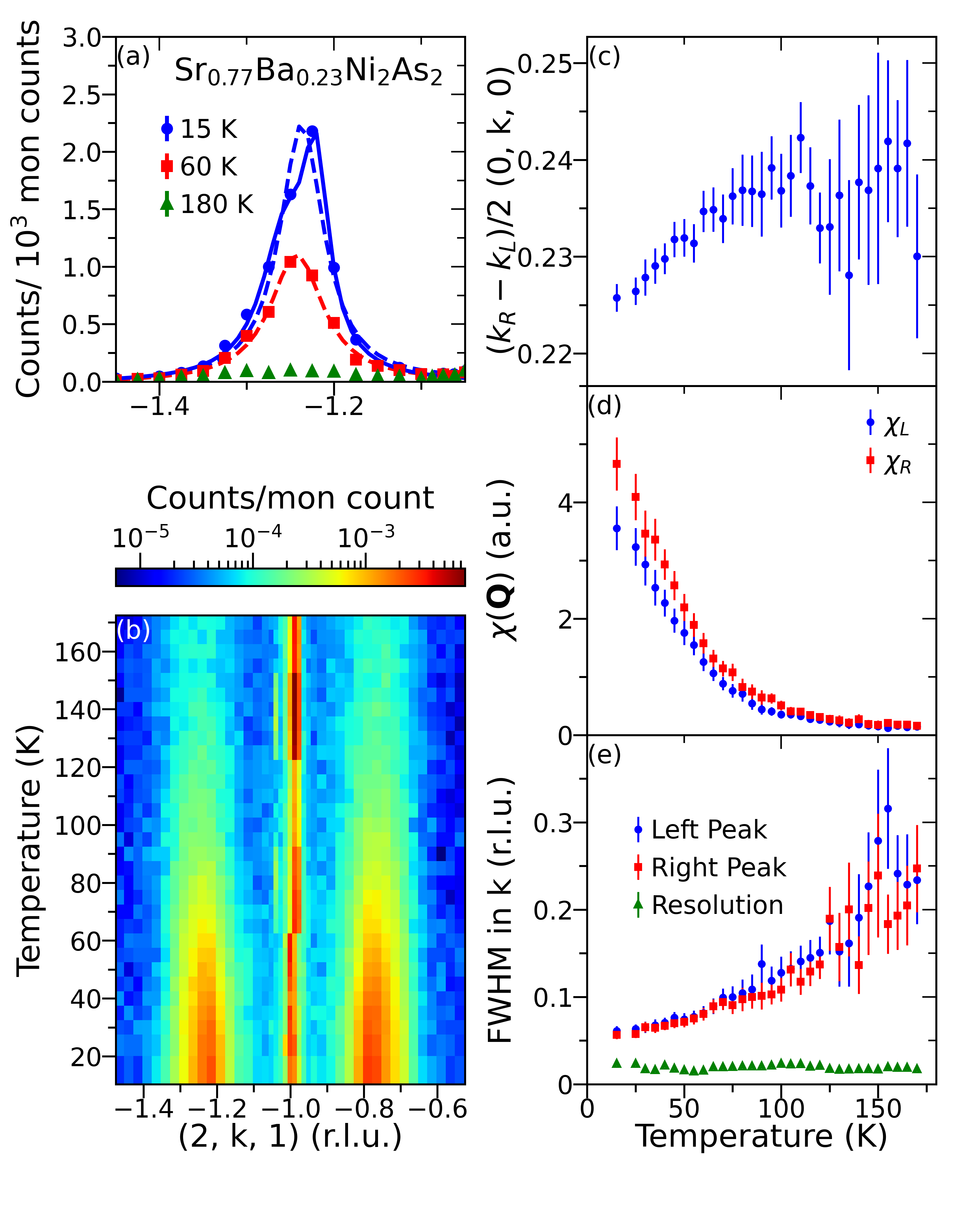}
    \centering
    \caption{The elastic X-ray scattering results and their fits are presented in panels (a),(b) and (c)-(e) respectively. The fitting function is a linear combination of single Lorentzian for the satellite peaks and a Gaussian for the central peak. (a) Incommensurate peaks observed at T~=~15, 60, and 180~K, respectively. The dashed lines are single Lorentzian fits, while the solid lines are fits using two Lorentzians. (b) Color plot for scans along the (2,~k,~1) direction at various temperatures ranging from 170 to 15~K. Data are presented on a logarithmic scale to visualize weak fluctuations at higher temperatures. The change in intensity of the central Bragg peak is due to differences in alignment at various temperatures. (c) Temperature dependence of the propagation vector for incommensurate charge density fluctuations. At higher temperatures, the peak position is not well determined due to relative weak scattering signal. (d) Temperature dependence of peak charge susceptibility \(\chi(\mathbf{Q})\). The blue and red data points represent fitting results for the left and right peaks, respectively. (e) Temperature dependence of FWHM for the left (blue), right (red), and central (green) peaks. The green data points show that the satellite peaks are much broader than the resolution-limited structural peaks.}
    \label{Fig_2}
\end{figure}

We begin our discussion by presenting elastic scans of Sr\(_{0.77}\)Ba\(_{0.23}\)Ni\(_{2}\)As\(_{2}\) along the (2,~k,~1) direction in Figs.~\ref{Fig_2}(a) and (b). We observed a pair of incommensurate peaks symmetric about (2,~0,~1) with a propagation vector \(\mathbf{q} = (0, 0.23, 0)\) at T~=~15~K. In the previous work, these satellite peaks were observed at a slightly different wave vector, \(\mathbf{q} = (0, 0.28, 0)\), and were attributed to an ICDW that emerges just before the onset of the triclinic phase~\cite{PhysRevLett.127.027602}. This triclinc phase is fully suppressed once the doping level crosses \( x_{c1} \), marking a first order phase transition. Our measurements, on the other hand, show that the incommensurate peaks not only survive the first-order quantum phase transition but also form at a high temperature similar to that in pristine BaNi\(_{2}\)As\(_{2}\). These features suggest that the incommensurate peaks are closely related to the tetragonal phase of Sr\(_{0.77}\)Ba\(_{0.23}\)Ni\(_{2}\)As\(_{2}\) and are not strongly correlated with the lattice dynamics that facilitate the triclinic phase. 

To understand the temperature dependence of the incommensurate peaks, we fit the lineshapes of elastic scans using a single Lorentzian function for the satellite peaks and Gaussian for the structural peaks at the center. The obtained fitting parameters are presented in Figs.~\ref{Fig_2}(c)-(e). The temperature dependence of the propagation vector shown in Fig.~\ref{Fig_2}(c) is continuous and does not lock onto any fractional wave vector throughout the entire temperature range, indicating its incommensurate nature. The fitted peak susceptibility \(\chi(\mathbf{Q})\), plotted in Figure~\ref{Fig_2}(d), exhibits a non-saturated behavior starting at approximately 150~K down to the lowest achievable temperature of 15~K. This is accompanied by a continuous peak sharpening, which is parameterized by the FWHM presented in Figure~\ref{Fig_2}(e). Considering that the incommensurate peaks are significantly broader than the resolution-limited structural peak at the center (represented by green triangles) throughout the entire temperature range, these observations suggest that short-range, quasi-elastic CDF persists down to 15~K.

We note that the linear extrapolated FWHM at absolute zero remains significantly larger than the instrumental resolution, suggesting a non-divergent behavior of the correlation length. This apparent conflict with a system near a QCP can be understood through the coupling between charged dopants and periodic lattice distortions via Friedel oscillations~\cite{PhysRevB.32.2449}. The dopants perturb the original charge density correlations, leading to secondary scattering close to the CDF wave vector. Such effects have been commonly observed~\cite{Mallayya2024,Yue2020}, most notably in cuprates~\cite{doi:10.1126/science.aav1315,PhysRevX.9.031042}. In the case of Sr\(_{0.77}\)Ba\(_{0.23}\)Ni\(_{2}\)As\(_{2}\), where the introduced charge potentials from dopants within the same periodic family are weak, the resulting secondary scattering is weaker compared to the main peak. This secondary scattering shares a similar temperature dependence with the primary CDF peak in terms of scattering intensity and occurs near the main peak. This proximity effectively broadens the CDF peak and introduces asymmetry into the peak profile at lower temperatures~\cite{PhysRevB.74.174102}. Indeed, at 15~K, we observed that the CDF peak profile becomes asymmetric and can be better fit with two Lorentzians, as shown in Fig.~\ref{Fig_2}(a). This peak asymmetry is not due to mosaic distribution, as evidenced by the satellite peaks at (2, -0.78, 1) and (2, -1.22, 1), which exhibit more intensity when \( \mathbf{Q} \) is further from the central Bragg peak at (2, -1, 1)~\cite{SM}. Alternatively, the broadening can be discussed in the language of statistical physics. The CDW in this system has the symmetry of a 3D XY model~\cite{Bruce_1978}. When a dopant is introduced, it couples to the CDW like a random field, effectively reducing the upper critical dimensional from 4 to 2~\cite{PhysRevLett.35.1399}. This reduction implies that in 3D, long-range order cannot be formed, which is consistent with our observations. Our discussion about these satellite peaks in this work remains within the framework of fitting a single Lorentzian, as a more detailed analysis of the peak profile would require higher sampling in scans than the current experimental data.

\begin{figure*}[t]
    \includegraphics[width=\textwidth]{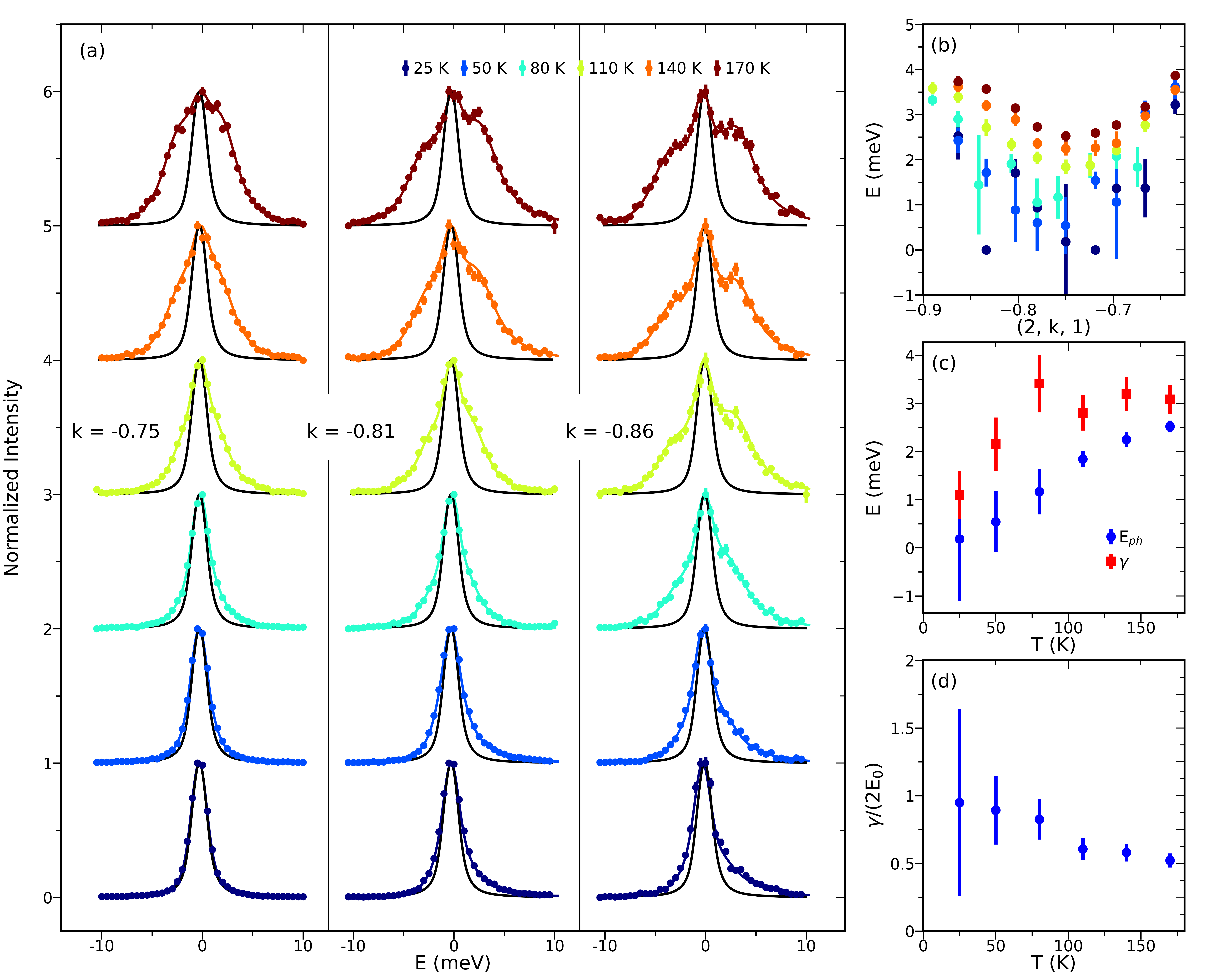}
    \centering
    \caption{The ICDW phonon results. (a) Energy scans at $\mathbf{Q}$ = (2,~-0.75,~1), (2,~-0.81,~1), and (2,~-0.86,~1). The measurements are taken at T = 25, 50, 80, 110, 140, and 170~K. The black solid line represents the instrument resolution. The spectra at each specific temperature are normalized to their maximum intensity to highlight the differences in line shape. (b) The phonon dispersion along the \((2, k, 1)\) direction was obtained using a DHO model at the temperatures mentioned in (a). The ICDW phonons start to become overdamped at 25~K, causing large uncertainties in the fitting parameters. (c) Temperature dependence of the phonon and line widths at $\mathbf{Q}$ = (2,~-0.75,~1). The phonon line width exhibits a linear relationship below 80 K, and the phonon becomes fully softened at \(T \approx 25\)~K. (d) Temperature dependence of damping ratio $\gamma/(2E_{0})$ at $\mathbf{Q}$ = (2,~-0.75,~1).}
    \label{Fig_3}
\end{figure*}

To study the low-energy CDFs, we measured the phonon spectra in the vicinity of the incommensurate peak location using inelastic X-ray scattering techniques. The phonon spectra at $\mathbf{Q}$ = (2,~-0.75,~1), which is close to the incommensurate peak location, are presented in Fig.~\ref{Fig_3}(a). The origin of the broad quasi-elastic line observed in Fig.~\ref{Fig_2} can be further inferred from the phonon spectra measured between 170~K and 25~K. At 25~K, the spectrum remains slightly broader than the instrumental resolution. This observation unambiguously evidences a continuous softening of the ICDW phonon down to the lowest achievable temperature.

A quantitative analysis of the spectra requires fitting the experimental data against the convolution of the instrument resolution with a damped harmonic oscillator (DHO) model~\cite{Baron2019}: 
\begin{equation}
    S (E) = \frac{A}{1-\exp(-E/k_{B}T)}\frac{\gamma E}{(E^{2}-E^{2}_{0})^2+(\gamma E)^2}
\end{equation}
where $A(\mathbf{q})$ is proportional to the phonon structure factor, $E_{0}(\mathbf{q})$ is the undamped ICDW phonon energy and $\gamma(\mathbf{q})$ is the phonon line width. When $\gamma/(2E_{0}) < 1$, the DHO model describes underdamped phonons whose energy can be determined as:
\begin{equation}
    E_{ph}=\sqrt{E_{0}^{2}-\gamma^{2}/4}
\end{equation}

The obtained ICDW phonon dispersion at 25, 50, 80, 110, 140 and 170~K are plotted in Fig.~\ref{Fig_3}(b). The phonon energy is well determined from spectra measured above 25~K, showing a dip at $\mathbf{Q}$ = (2,~-0.75,~1). At 25~K, the fitting results shows the ICDW phonon are either in or close to the overdamped region ($\gamma$/(2$E_{0}$) $>$ 1) where lattice relaxations rather than atomic oscillations are present. This results in the phonon spectral weight overlapping with the elastic signal, creating technical difficulties in accurately fitting the spectra using a DHO model. Nevertheless, our fit results with finite phonon energy over a large temperature range confirm that our observation of incommensurate peaks originates from dynamical fluctuations, which naturally account for their short-ranged nature. The phonon dispersion within \(\mathbf{Q} = (2, -0.75, 1)\) and \((2, -0.9, 1)\) is approximately linear at \(T = 170\) K. This allows us to estimate the ICDW phonon energy at the gamma point using linear extrapolation, yielding an estimated optical phonon energy at the Brillouin zone center \(E_{ph,\Gamma} \approx 5.0 \text{ meV}\). The obtained phonon energy is consistent with the energy of the \(E_{g}\) phonon mode observed in recent Raman experiments at T~=~160~K, where significant splitting is found when the ICDW develops in BaNi$_{2}$(As, P)$_{2}$~\cite{Yao2022}. 

The temperature dependence of the phonon line width and damping ratio at $\mathbf{Q}$ = (2,~-0.75,~1) is plotted in Figs.~\ref{Fig_3}(c) and (d), respectively. Notably, significant damping of the phonons already exists at 170~K with \(\gamma \approx 3 \, \text{meV}\), which remains temperature-independent until 80~K. At 80~K the phonon line width starts to drop linearly, a behavior also observed prior to the formation of the ICDW in pristine BaNi\(_2\)As\(_2\)\cite{PhysRevB.107.L041113}. This linear decay of line width is also observed in magnons during the formation of antiferromagnet Fermi liquids. In antiferromagnetic Fermi liquids, for critical fluctuations near the ordering vector $\mathbf{Q}_{\mathrm{ICDW}}$ where \(|\mathbf{Q} - \mathbf{Q}_{\mathrm{ICDW}}|\) is an order of magnitude smaller than the inverse correlation length, the temperature dependence of the phonon energy line widths can be simplified as follow:
\begin{equation}
    \gamma=\gamma_{0}(T+\theta)
\end{equation}
where $\theta$ is the Curie-Weiss temperature. The linear fit using line widths at T = 25, 50 and 80~K gives $\theta$~=~0 (10)K, consistent with the ordering temperature of a system at the vicinity of a QCP~\cite{moriya1985spin}. Due to the large uncertainties in the fitted phonon line widths and the lack of sufficient data near absolute zero, a finite phonon energy width corresponding to a finite correlation length (Fig.~\ref{Fig_2} (e)) at 0 K, may still be possible.

\begin{figure}[t]
    \includegraphics[width=\columnwidth]{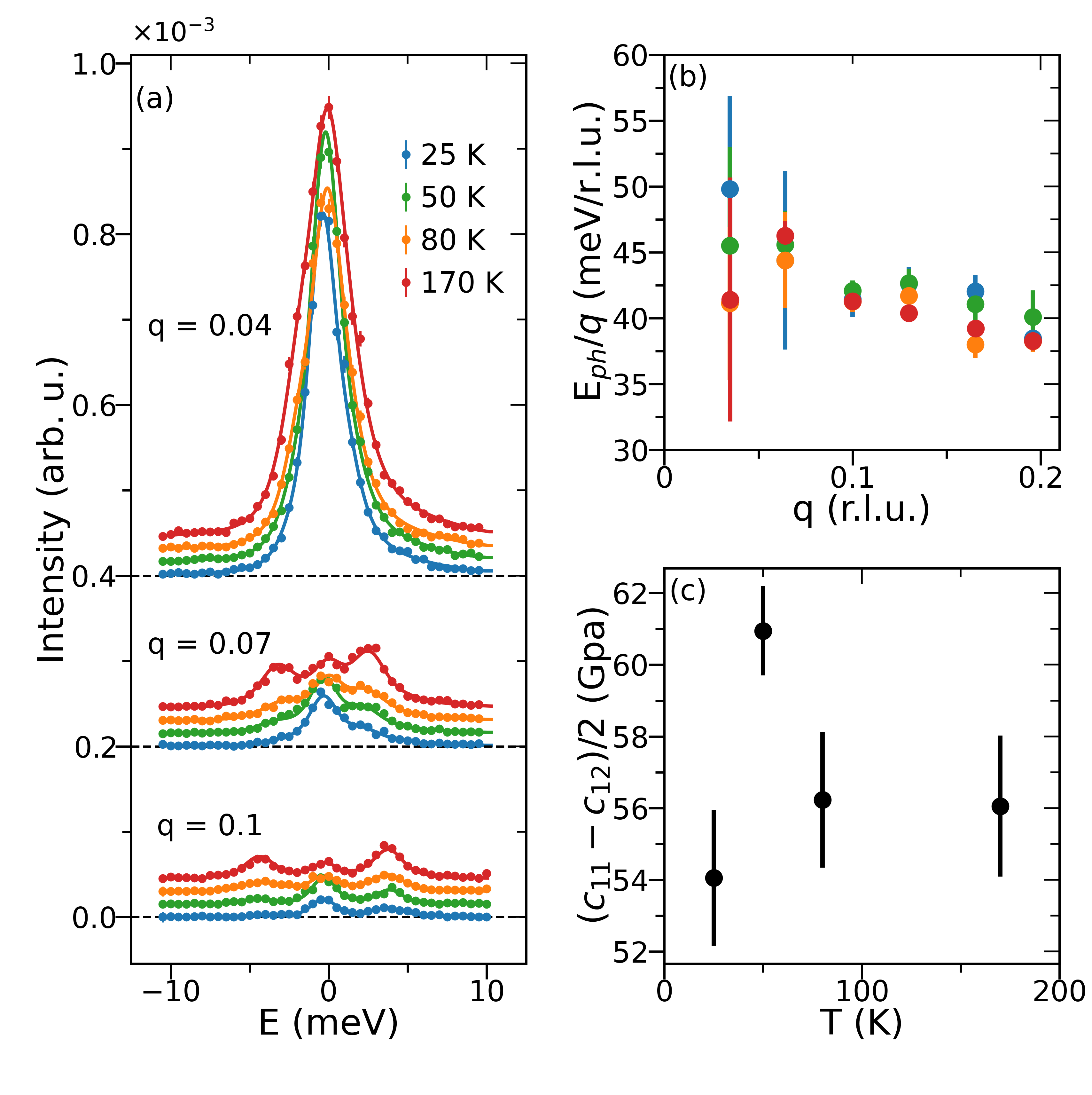}
    \centering
    \caption{The IPTA phonon results. (a) Energy scans at $\mathbf{Q}$ = (3+$q$, -3+$q$, 0) with $q$ = 0.04, 0.07 and 0.1~r.l.u. The measurements are taken at T~=~25, 50, 80 and 170~K. The solid line represents the fitting result using a DHO model convoluted with the instrument resolution. (b) The phonon dispersion along the $\mathbf{Q}$ = (3+$q$, -3+$q$, 0) direction was obtained using a DHO model at the temperatures mentioned in (a). The IPTA close to the Brillouin zone center does not experience significant softening over the whole temperature region. (c) Temperature dependence of elastic modulus $(c_{11}-c_{12}/2)$. The value is obtained via fitting the dispersion in (b) against a phonon dispersion $E_{ph}=A\sin(\pi q/2)/(\pi/2)$. The change of the value is not significant compared to the error bars.}
    \label{Fig_4}
\end{figure}

For low-energy nematic fluctuations, previous studies in electron- and hole-doped BaFe\(_2\)As\(_2\) have shown that they can result in the softening of the in-plane transverse acoustic (IPTA) phonon. Specifically, the dispersion of the IPTA phonon near the Brillouin zone center can be precisely described by mean-field theory in the weak coupling scenario using the following formula~\cite{PhysRevLett.126.107001}:

\begin{equation}
    \frac{E_{\xi}(k)}{E_{\xi=0}(k)}=\sqrt{\frac{1+\xi^{2}k^{2}}{1+\xi^{2}(k^{2}+r)}},
\end{equation}

\noindent where \(\xi\) is the nematic correlation length and \(r\) is a temperature-independent parameter. Yet, unlike iron-based superconductors, the elastoresistivity peak is present only in the B\(_{1g}\) channel for BaNi\(_2\)As\(_2\). The coupled elastic modulus is then \((c_{11}-c_{12})/2\), which corresponds to the sound velocity of the IPTA phonon along the \(\langle 110 \rangle\) directions. Fig.~\ref{Fig_4}(a) presents the IPTA phonon spectra measured at \((3+q, -3+q, 0)\), where \(q\) is selected as 0.04, 0.07, and 0.1~r.l.u. in order to capture potential IPTA softening in the vicinity of the gamma point. However, we found no significant change in the line shape of the spectra when T~$<$~80~K. To further investigate the temperature dependence of IPTA phonons, their phonon dispersion at 25, 50, 80 and 170~K, extracted using a DHO model, are displayed in Fig.~\ref{Fig_4}(b). Contrary to the iron-based superconductors, the dispersion of IPTA phonon in Sr$_{0.77}$Ba$_{0.23}$Ni$_{2}$As$_{2}$ does not change and even slightly hardens upon cooling, which suggests a minor effect of nematic fluctuations on the lattice. This can also be inferred from the absence of a divergence of the fitted elastic modulus \((c_{11}-c_{12})/2\), which is presented in Fig.~\ref{Fig_4}(c).

The absence of IPTA phonon softening at \(\mathbf{Q} = 0\), is consistent with X-ray diffraction results presented in Fig.~\ref{Fig_2}, where no broadening or splitting of the central Bragg peak was observed over the whole temperature region. Our results show the previously reported electronic nematicity in Sr$_{x}$Ba$_{1-x}$Ni$_{2}$As$_{2}$ is not lattice driven. Notably, the E$_{g}$ phonon mode that condenses together with the ICDW splits upon the formation of the ICDW, and density perturbation theory predicts that the phonon dispersion for this E$_{g}$ phonon is anisotripic along the (100) and (010) directions~\cite{PhysRevLett.129.247602}, suggesting a 4-fold symmetry breaking in the ICDW phase. Such 4-fold summery breaking can happen at a finite \textbf{q} with a non-zero orbital polarization:
\begin{equation}
\hat{O}_{\mathbf{q}}=\sum_{i,\sigma}\mathrm{e}^{j\mathbf{q}\mathbf{r}_{i}}(d_{i\sigma,xz}^{\dagger}d_{i\sigma,xz}-d_{i\sigma,yz}^{\dagger}d_{i\sigma,yz})\neq 0
\end{equation}
This will cause the ICDW to become unidirectional and potentially exhibit an orbital character, which might be detectable through resonant X-ray scattering using different polarizations, similar to what has been observed in cuprates~\cite{doi:10.1126/sciadv.aay0345}. As for the superconductivity, the theoretical framework of the iron pniticides points out that charge fluctuations will give raise to a conventional s-wave superconductor~\cite{annurev:/content/journals/10.1146/annurev-conmatphys-020911-125055}. Considering that the phonon dispersion of Sr\(_{0.77}\)Ba\(_{0.23}\)Ni\(_{2}\)As\(_{2}\) can be described using the formula typically applied to the antiferromagnetic Fermi liquid model, which also characterizes the temperature dependence of magnetic fluctuations above the superconducting temperature in optimally doped BaFe\(_{1.85}\)Co\(_{0.15}\)As\(_{2}\)~\cite{Inosov2010}, it would be intriguing to investigate whether there is a resonant feature in the spectrum that couples to the ICDW phonon below the superconducting transition temperature of 3.3 K. Such a feature could reveal a deeper connection between the ICDW fluctuations and the emergence of superconductivity, potentially offering insight into the role of charge density wave fluctuations in the superconducting state. 

In summary, our work revealed intense quasi-elastic scattering and a pronounced softening of the ICDW phonon in Sr\(_{0.77}\)Ba\(_{0.23}\)Ni\(_{2}\)As\(_{2}\). A quantitative analysis of the softened phonon indicates that ICDW is in the vicinity of the proposed quantum critical point. Regarding the orbital degrees of freedom, we observed no softening of the IPTA phonon, suggesting that the electronic nematicity is not lattice-driven. Our findings imply that charge density fluctuations may play a crucial role in the enhancement of superconductivity at the proposed quantum critical point of Sr\(_{x}\)Ba\(_{1-x}\)Ni\(_{2}\)As\(_{2}\).

\begin{acknowledgements}
The work at UC Berkeley and LBNL was funded by the U.S. Department of Energy, Office of Science, Office of Basic Energy Sciences, Material Sciences and Engineering Division under Contract No. DE-AC02-05-CH11231 (Quantum Materials program KC2202). The synchrotron radiation experiments were performed at BL35XU of SPring-8 with the approval of the Japan Synchrotron Radiation Research Institute (JASRI) under Proposal No. 2023B1149. The work at Zhejiang University was supported by the Pioneer and Leading Goose R\&D Program of Zhejiang (Grant No. 2022SDXHDX0005), the National Key R\&D Program of China (Grant No. 2022YFA1402200), the Key R\&D Program of Zhejiang Province, China (Grant No. 2021C01002), and the National Natural Science Foundation of China (Grant No. 12350710785).
\end{acknowledgements}

\bibliography{apssamp}

\end{document}